\documentclass[runningheads]{svmult}
\usepackage{makeidx}   
\usepackage{graphicx}  
\usepackage{subeqnar}  
\usepackage{multicol}  
\usepackage{cropmark} 
\usepackage{physprbb}  
\def\Journal#1#2#3#4{{#1} {\bf #2}, #3 (#4)}

\def\NPB{{\em Nucl. Phys.} B}
\def\PLB{{\em Phys. Lett.}  B}
\def\PRL{\em Phys. Rev. Lett.}
\def\PRD{{\em Phys. Rev.} D}
\def\ZPC{{\em Z. Phys.} C}
\def\beq{\begin{equation}}
\def\eeq{\end{equation}}
\def\beqn{\begin{eqnarray}}
\def\eeqn{\end{eqnarray}}

\begin{document}
\title*{Twenty Years of SUGRA}
\toctitle{   Twenty Years of SUGRA     }
%
%
\titlerunning{Twenty Years of SUGRA}
%
\author{Pran Nath
}
\authorrunning{Pran Nath}
%
%
\institute{Department of Physics, Northeastern University, 
Boston, Massachusetts,  02115, USA}

\maketitle              

\begin{abstract}
A brief review is given of the developments of mSUGRA and its
extensions since the formulation of these  models in 1982.
Future directions and prospects are also discussed.
\end{abstract}

\section{Introduction}
Supersymmetry provides a technical solution to the so called 
gauge hierarchy problem in the form of a no renormalization 
theorem\cite{grs} which makes it an attractive candidate for
 model building. The main hurdle 
in the development of realistic models in the early days was
the difficulty of breaking supersymmetry\cite{ww} in a phenomenologically
viable manner\cite{georgi}. The resolution of this problem arises in supergravity framework.
In this paper we briefly summarize the developments over the 
last 20 years since the formulation of the minimal supergravity grand 
unified model\cite{msugra,applied} (mSUGRA) and its extensions 
including nonuniversalities to which we give the generic name: SUGRA models.
 The formulations of SUGRA models are based on techniques of applied 
supergravity where one couples gauge fields with matter fields ($z_i$)
and then couples the combined system to $N=1$ 
supergravity\cite{sugracouplings,applied}. The coupled theory depends
on three arbitrary functions: the gauge kinetic energy function
$f_{\alpha\beta}(z_i,z_i^{\dagger})$, the Kahler potential
$K(z_i,z_i^{\dagger})$, and the superpotential $W(z_i)$. 
This allows one to construct
grand unified models based on supergravity\cite{cgrand}. 
The central assumption of SUGRA models is that supersymmetry is broken
in a so  called hidden sector and the breaking is communicated by
gravitational interactions to the visible sector where quarks, leptons
and the Higgs fields reside\cite{msugra}. The theory intrinsically contains
large mass  scales which include the Planck scale and the 
grand unification scale. It is then shown that the  low energy
theory that results after integration over the Planck scale and the GUT scale 
is free of both these high scales, i.e., the Planck scale and the
GUT scale. The absence of the Planck scale from low energy
theory was shown in the work of Chamseddine etal and Barbieri etal
in Ref.\cite{msugra} while the cancellation of the grand unification scale
was shown in the work of Chamseddine etal and Hall eta  in Ref.\cite{msugra}.
 For the case of minimal supergravity unification which uses the 
 flat Kahler potential $(K=\sum z_iz_i^{\dagger})$\cite{msugra} the low energy theory results in just
four soft breaking parameters which are  the universal scalar mass
$m_0$, the universal gaugino mass $m_{\frac{1}{2}}$, the universal
trilinear coupling $ A_0$ and the universal bilinear couplings $B_0$.
In addition, the low energy theory contains a Higgs mixing parameter 
$\mu_0$ which can arise in a variety of ways but its size is
typically of the soft breaking scale\cite{msugra,applied,soni,muterm}.
The universality of the soft parameters holds at 
unification scale and below this scale the soft parameters and the parameter 
$\mu$ evolve according to the renormalization group equations governing the
gauge and yukawa couplings\cite{einhorn} and  the soft parameters\cite{inoue}. 
In the development of mSUGRA models it was assumed that SUSY breaks
in the hidden sector via some scalar fields developing vacuum 
expectation values\cite{polonyi}. At the more fundamental level
the breaking may arise via gaugino condensation with 
$<\lambda  \gamma^0 \lambda>\neq 0$\cite{nilles1}.
 However, this phenomenon requires
non-perturbative effects to occur and generally it is difficult to obtain
explicit models where a satisfactory solution is achieved.
Returning to 
SUGRA models, they have played a dominant role in the development of SUSY 
phenomenology and some of the early works are contained in 
Ref.\cite{earlypheno}.

A remarkable aspect of SUGRA models is that they lead to the breaking
of the electroweak symmetry\cite{msugra} which is something that is
rather adhoc in the standard model.  Further, an attractive mechanism for
this is via renormalization group effects using renormalization
group evolution\cite{inoue}. 
The radiative electroweak symmetry breaking solutions 
must be subject to the constraints of color and charge 
conservation\cite{frere}. Under these constraints one minimizes the
effective potential in the vicinity of the electroweak scale
which leads to constraints on the vacuum expectation value of
the Higgs fields. In the minimal supersymmetric standard model (MSSM)
one has two Higgs fields, $H_i$ (i=1,2), which leads to two
constraints which arise from the extremization conditions corresponding
to these fields.  
One of these constraints can be used to eliminate $\mu$
(the value of $\mu_0$ at the electroweak scale) 
except for its sign while the second one allows one to eliminate
the soft parameter $B_0$ in favor of $\tan\beta =<H_2>/<H_1>$, 
where $H_2$ gives
mass to the up quark and $<H_1>$ gives mass to the down quark and the 
lepton. Thus the low energy theory  can be described
by only four parameters, i.e. the parameters $m_0$, $m_{\frac{1}{2}}$, 
$A_0$, $\tan\beta$, and sign ($\mu$). 
This is to be contrasted with the large number of soft
parameters one can add in MSSM\cite{Dimopoulos:1995ju} 
consistent with the cancellation of the quadratic 
divergences\cite{grisaru}.
Historically local supersymmetry first arose in the form of
supergravity in superspace called gauge 
supersymmetry\cite{Nath:1975nj,Arnowitt:1975xg}. 
This formulation was a direct extension of 
Einstein gravity to superspace. The formulation of local supersymmetry
directly in ordinary space was given in 
Refs.\cite{Freedman:1976xh,Deser:1976eh}. 

The outline of the rest of the paper is as follows: In section 2 
we discuss some of the signatures of SUGRA models which include 
a discussion of the phenomenon of scaling. 
We discuss the trileptonic signal and also emphasize the
importance of the process $B^0_{s,d}\rightarrow \mu^+\mu^-$ as
an important signal of SUSY and SUGRA for the case of large
$\tan\beta$. In Sec.3 we discuss the current situation regarding
the implications of the Brookhaven experiment 
on $a_{\mu}=(g_{\mu}-2)/2$ for 
SUGRA models. It is argued that a  SUSY contribution to
$a_{\mu}$ of size $O(10^{-9})$ or larger implies upper limits for sparticle
masses which should be accessible at the Large Hadron Collider 
(LHC). In Sec.4 we discuss the implications of CP phases arising from the
soft breaking parameters becoming complex. In this section
we also discuss the constraints on CP phases from the electric
dipole moment limits that exist for the electron, the neutron
and for the atomic edms such as $Hg^{199}$. Implications of large
CP phases on low energy phenomena are also discussed. Of special
interest are the phenomena involving  the mixings of CP even
and CP odd Higgs states induced by the CP phases via loop
corrections to the Higgs masses. These mixings can lead to
interesting effects in SUSY phenomena at colliders. In Sec.4
we also discuss the effects of CP phases on supersymmetric
corrections to the $b$ quark mass and for the $\tau$ lepton mass,
and the modifications of the Higgs decays 
$h\rightarrow b\bar b, \tau\bar\tau, c\bar c$ due to CP phases.
An accurate measurements of these decays may reveal the effects
of supersymmetry and of CP phases. In Sec.5 we discuss the 
hyperbolic branch (HB) of the radiative breaking in SUGRA models.
This branch is characterized by the property that the soft
parameters $m_0$ and $m_{1/2}$ lie on the surface of a 
hyperbola for fixed $\mu$ and thus can get large while $\mu$
remains fixed and small. It is also discussed in Sec.5 that
the focus point (FP) region is a part of the hyperbolic branch 
corresponding to relatively low values of $m_{1/2}$.
 The hyperbolic branch naturally leads to heavy
squarks and gluinos but it is shown in Sec.6 that this branch
may still allow for sufficient relic density to be consistent with 
the current astrophysical data. In Sec.7 the current status of
SUGRA GUTs is discussed. Also discussed is the relationship of
SUGRA to strings. Conclusions are given in Sec.8.
In this paper we do not 
discuss other developments such as the no scale supergravity\cite{lahanas}
and the gauge mediated breaking of supersymmetry\cite{rattazzi}.
Indeed the literature  following the development of SUGRA models
is enormous and it is physically not possible to do justice to 
reviewing it in a conference talk. Thus although the bibliography
looks extensive there is no claim it is complete
as it is only a fraction of the existing literature. Further, we
 limit ourselves in this talk to only a few topics of current 
 interest and many other interesting topics within SUGRA are not 
 discussed.

\section{SUGRA Signatures}
An important hint for supersymmetry comes from the LEP data in terms of
precise values for the gauge couplings constants and the fact that
unification occurs within SUSY/SUGRA unified 
models\cite{drw,couplings,gravi-couplings}.
We note that the threshold corrections from the sparticle masses
enter crucially in the gauge coupling unification and provide an
indirect support for SUGRA. 
The analyses involve evolution of gauge and Yukawa couplings 
from the GUT  scale to low energy using one and two loop
renormalization group equations of the gauge and Yukawa
couplings\cite{einhorn,marie} and of the soft 
parameters\cite{inoue,martin}. Further, more accurate determinations
of the sparticle masses at low energy require that the minimization
 of the effective potential include loop 
 corrections\cite{coleman,gamberini,anloop}. In the Higgs sector
 such loop corrections turn out to be crucial. Thus in
 mSUGRA one has the tree relation that the lightest Higgs
mass should lie below the Z boson mass\cite{applied}. 
However, this relation is modified by loop corrections which 
lift the lightest Higgs mass significantly above $M_Z$\cite{higgsloop}.
The parameter space of SUGRA model is limited by 
experimental and theoretical constraints. One of these
is the constraint of the flavor changing neutral current
process $b\rightarrow s+\gamma$ which
limits the parameter space depending on the $\mu$ sign\cite{bsgamma}.

An interesting  phenomenon exhibited by renormalization 
group analyses is that of scaling\cite{scaling}. It 
arises in regions of the parameter space where $\mu>> M_Z$ 
(and $m_0$, $m_{\tilde g}$ $\leq 1$ TeV)
and leads to interesting relation on the 
gaugino masses of the type  $m_{\chi_1^0}\simeq 0.5 m_{\chi_2^0}$,
$m_{\chi_2^0}\simeq  m_{\tilde g}/3$,  $m_{\chi_2^0}\simeq m_{\chi_1^{\pm}}$,
$m_{\chi_3}\simeq m_{\chi_4^0}\simeq \mu$ as well as interesting
relations among the other sparticle masses. 
Many of the mSUGRA mass relations can be appropriately expressed as sum rules
which can be put to test when one has a precise measurement
of the sparticle masses\cite{scaling,ramond}. One consequence of
the RG analysis is that the lightest neutralino turns out to be
the lightest supersymmetric particle (lsp) over most of the
parameter space of the model. Further, in the region where scaling 
holds one finds that the lightest neutralino is in fact mostly
a Bino\cite{scaling,roberts}. This result has important implications
for supersymmetric dark matter in SUGRA models with R parity conservation.
Further, the R parity conservation constraint leads to interesting missing
energy signals in sparticles decays.  
One such signal is the trileptonic signal. It was noted early
in the investigation of SUGRA models that the $W^{\pm}$ decay into a
chargino and the second lightest neutralino, i.e., 
$\chi_1^{\pm}+\chi_2^0$ with the subsequent decays of the $\chi_1^{\pm}$
and  $\chi_2^0$ can lead to a clean trileptonic signal\cite{earlypheno}
 and further work was carried out in Ref.\cite{Baer:1985at}.
 In Ref.\cite{Nath:sw} it was observed that the decays from an 
 off-shell W can extend very significantly the potential for the 
 discovery of the chargino $\chi_1^{\pm}$.
 A recent analysis of this process can be found in Ref.\cite{Baer:2003dj}.
 Another process which
 has been the subject of recent studies is the decay
 $B^0_{s}\rightarrow \mu^+\mu^-$.
  In the standard model the branching ratio for this process is
  rather small, i.e., 
  $Br(\bar B^0_s\rightarrow \mu^+\mu^-)=(2.1\pm 1.4)\times 10^{-9}$ 
  (for $V_{ts}=0.04\pm 0.002$) and beyond the reach of experiment
  in the near future. Thus the current experimental limit for
  this process is  
$Br(\bar B^0_s\rightarrow \mu^+\mu^-)<2.6\times 10^{-6}$
and one estimates that a branching ratio down to the level of
$10^{-8}$ may be achievable at RUNII of the Tevatron. However, 
a check on the   standard model branching ratio 
$Br(\bar B^0_s\rightarrow \mu^+\mu^-)$ 
would not be feasible even with this enhanced 
sensitivity.   In supersymmetry the 
so called counterterm diagram is proportional to $\tan^6\beta$
for large $\tan\beta$\cite{gaur}. Detailed numerical analyses
show that indeed one expects a big enhancement of order $10^2$ in
some parts of the parameter space in SUGRA 
models\cite{Dedes:2002zx,Arnowitt:2002cq,baek,miztata,huang,Ibrahim:2002fx}.
Thus the process $Br(\bar B^0_s\rightarrow \mu^+\mu^-)$  is a
strong indicator of supersymmetry and in fact an observation of this
process will be a pointer to the existence of sparticles even 
before the sparticles are seen. There are a variety of other  
 signals which have been discussed in the literature and the 
 reader is directed to  Ref.\cite{sugra} for a more comprehensive
 survey, to Refs.\cite{Baer:2003dj,Baer:2003wx,Baer:2003yh} for
 more recent constraints on mSUGRA, to
 Refs.\cite{arnowittsugra20,baersugra20,tatasugra20} 
 for more recent surveys, and to Refs.\cite{paigesugra20,villasugra20}
 for the disovery potential of SUSY/SUGRA at ATLAS and CMS.

\section{$g_{\mu}-2$ and SUGRA}
$g_{\mu}-2$ is one of the most sensitively determined quantities
in all of physics. The precision of this determination has
further increased due to the recent Brookhaven 
experiment\cite{brown,bnl}. The standard model correction consists 
of several parts: the qed correction, the hadronic correction
and the electroweak correction\cite{czar1}. Of these the 
qed correction and the electroweak correction are reliably
determined and the largest source of error in the theoretical 
analysis arises from the hadronic correction. 
The hadronic correction consists of the leading order (LO) and the nonleading 
order hadronic (NLO) contributions and the light-by-light hadronic
 contribution. The NLO correction is well understood. However, the LO 
 correction has been the subject of much 
 scrutiny\cite{davier,hagiwara}.
Similarly, the light-by-light hadronic correction has seen 
a flip in its sign and is still a subject of some 
debate\cite{knecht,hkrevised,Blokland,Bijnens2,Ramsey}.
A very recent estimate of the difference between experiment
and theory  gives\cite{hagiwara} $\Delta a_{\mu}$=
 $(a_{\mu}^{exp}-a_{\mu}^{SM})\sim (33\pm 10)\times 10^{-10}$
which is about  a $3\sigma$ effect. However, this difference
is likely to  change with more accurate determinations of the 
LO hadronic correction, and with more data expected from Brookhaven. 
Now it was predicted already nearly twenty years ago that the 
supersymmetric electroweak correction should be of the same size
as the standard model electroweak correction\cite{yuan} and that 
any experiment that tests the standard model electroweak correction
will in fact also test the supersymmetric electroweak correction\cite{yuan}.
Further, it was later pointed out that the sign of susy correction
to $g_{\mu}-2$ is the sign of $\mu$ in a large part of the 
parameter space\cite{lopez,chatto}. An important issue concerns if 
extra dimensions might provide a strong background to the 
supersymmetric effects. That is to say if the corrections to 
$g_{\mu}-2$ from the exchange of Kaluza-Klein states might produce
a large enough correction which might mask the supersymmetric
effect. This analysis was carried out in Ref.\cite{ny2}.
The analysis showed that using the current lower limits on
the size of extra dimensions one finds that the corrections
from the extra dimensions do not produce a serious background 
to the supersymmetric correction.  Thus after the results of the
Brookhaven experiment in the year 2001\cite{brown} which showed
a $2.6\sigma$ effect there was a lot of 
theoretical activity to understand the implications of the 
results\cite{chatto2001} in the context of supersymmetry. 
One of the major consequences that emerged from these analyses
was the result that the Brookhaven result implied
the existence of upper limits on sparticle masses which
appeared to lie within reach of the LHC. The Brookhaven
result of 2002\cite{bnl} is essentially consistent with its
previous determination. However, in the meantime the theoretical
evaluations  of $\Delta a_{\mu}^{SM}$ have changed due to 
reevaluations of the leading order hadronic correction and the 
light-by -light hadronic correction. While the new evaluations
still indicate a significant effect\cite{hagiwara}, the situation
is in a state of flux since the issue of hadronic corrections
to $a_{\mu}^{SM}$ is not fully settled and there is
more data to come from Brookhaven.

\section{CP Phases}
The minimal supergravity model mSUGRA can be extended to have
two phases which can be taken to be the phase of $\mu_0$ and
the phase of the trilinear coupling $ A_0$. The main problem
encountered with the inclusion of phases is that there are severe 
experimental constraints on them, e.g. from the electron 
edm\cite{eedm} and from the neutron edm\cite{nedm}.
Theoretically there are many contributions to the electric 
dipole moment of an elementary particle.  Thus while for the
electron the contribution to the electric dipole moment arises
from the electric dipole operator, for the quarks it arises from
 the electric dipole operator, from the chromo electric
dipole operator and from the dimension six operator\cite{weinbergdim6}.
These contributions are computed at the weak scale and evolved
down to the low scales where the experimental measurements of
these are given. 
In extracting contributions from the chromo electric
dipole operator and from the dimension six operator one  uses 
the so called naive dimensional analysis\cite{manohar}.
 Now typically phases O(1) tend to give  edm contributions 
which are already in contradiction with current 
experiment\cite{eedm,nedm}.
There are a variety of ways that
have been discussed in the literature for controlling these
edms. These include fine tuning to make the phases small\cite{ellis},
suppression of the edms by heavy masses\cite{na}, suppression
of phases in a class of Left-Right symmetric models\cite{bdm2},
and suppression by the cancellation mechanism\cite{incancel,inbrane}.
Additional ways for the suppression include putting the phases
in the third generation\cite{chang}. Further, the experimental
limits on the atomic edms arising from Schiff moment and specifically 
the experimental atomic edm of
$Hg^{199}$\cite{atomic} also impose important constraints on
model building\cite{olive,inhg199}.
Since a broad class of SUGRA and string models constrain
soft breaking  parameters with large phases, the cancellation
mechanism is specially suited for these scenarios. We explain,
therefore, in some detail how the cancellation mechanism by which 
the edms are reduced works.
 Consider, for example, the electric
dipole moment of the electron which receives contributions only 
from the electric dipole operator. However, the supersymmetric contribution
to this operator includes contributions from the exchange of two 
charginos and four neutralinos. In certain regions of the parameter
space these contributions have opposite signs and naturally cancel
reducing the edm below the experimental limit.
 For the quarks the situation is more complex.
As mentioned above the quark edm receives contributions from the 
electric dipole, the chromoelectric dipole and the purely 
gluonic dimension six operator. 
Here the electric dipole operator and the chromoelectric dipole
operators receive contributions from the exchange of the gluino,
the charginos, and the neurtralinos and thus there are even 
more possibilities for cancellations. Similarly for the 
dimension six operators  one has 
contributions arising from the exchange of stops and sbottoms.
In addition to the cancellations that can occur within 
each individual operator from gluino, chargino and neutralino
exchanges, one has an additional  possibility for cancellations 
for the quark case
not available for the electron electric dipole moment, i.e., one
may have cross cancellations among the electric dipole, the 
chromoelectric dipole, and the purely gluonic dimension six 
operators. 

 Such cancellations are further facilitated in the  
nonuniversal SUGRA model (nSUGRA). In  nSUGRA one may give
nonuniversal gaugino masses as well as nonuniversalities 
in the Higgs sector and in the third generation sector consistent with
FCNC. We focus here on the gaugino sector. In this case one may
have independent gaugino masses for  the $SU(3)$, $SU(2)$ and 
$U(1)$ sector gaugino masses $\tilde m_i$  so that 
$\tilde m_i= |\tilde m_i| e^{i\xi_i}$ (i=1,2,3). We note in passing
that the physical quantities such as the edms depend only on certain
combinations of phases which have been classified in Ref.\cite{inmssm}.
Even so the appearance of a larger number of phases allows a larger
region of the parameter  space for the cancellation mechanism to
operate. The central point of the cancellation mechanism and of 
other mechanisms is that they allow the phases to be large and 
one can still satisfy the edm constraints. Now if the phases are large
they will affect a variety of low energy phenomena.
These include effects on Higgs phenomenology\cite{pilaftsis,inhiggs},
on sparticle phenomenology\cite{kane,zerwas}, on flavor
and B physics\cite{masiero1,masisugra20},  on 
$B^0_{s,d}\rightarrow \mu^+\mu^-$\cite{Ibrahim:2002fx},  
  on $g_{\mu}-2$\cite{Ibrahim:1999aj}, and 
  on proton decay\cite{Ibrahim:2000tx}.  CP phases also affect
  loop corrections to the $b$ quark mass and the $\tau$ 
  lepton mass\cite{Ibrahim:2003ca}.
  It is known that the SUSY effects can produce large corrections
  to the b quark mass for large $\tan\beta$\cite{susybtmass}.
  It was found in Ref.\cite{Ibrahim:2003ca} that CP effects on
  these can also be large. Similarly supersymmetric effects 
  produce important corrections to the Higgs decays to 
  $b\bar b$, $\tau \bar \tau$ and $c\bar c$\cite{babu1998}.
   Here also one finds that CP phases can produce large corrections
   to these decays of the Higgs bosons\cite{Ibrahim:2003jm}.

\section{Hyperbolic Branch/Focus Point (HB/FP)}
It is now known that there are two branches
to the radiative breaking of the electroweak symmetry, an
ellipsoidal branch and a hyperbolic branch\cite{ccn}. 
 These arise due to two solutions of $\mu$ using radiative 
 symmetry breaking equation that determines $\mu$, i.e., 
	$C_1m_0^2+C_3m'^2_{1/2}+C_2'A_0^2+\Delta \mu^2_{loop}$=
	$\mu^2$+$\frac{1}{2}M_Z^2$,
where $m_{1/2}'=m_{1/2}+\frac{1}{2}A_0C_4/C_3$, and $C_1$ etc
	are determined purely in terms of gauge and Yukawa couplings
	and $\Delta\mu^2$ are the loop corrections. These loop
	corrections play an important role in the analysis. For 
	small to moderate values of $\tan\beta$ the loop corrections
	 are relatively small. Also from the renormalization group
	 analysis one finds that the co-efficients $C_2'$, $C_3$
	 are positive. In these cases the scale dependence of 
	 $C_1$ is relatively small and one finds $C_1>0$ for
	 a range of scales $Q$ where the radiative electroweak
	 symmetry breaking is realized. In this case one finds that
	 the soft  parameters for fixed $\mu$ lie on the surface 
	 of an ellipsoid.  Now for larger values of $\tan\beta$,
	 i.e., typically $\tan\beta > 5$  one finds that the
	 loop corrections to $\mu$ become large. Further, for this
	 case one also finds a rather significant variation in this
	 correction with the scale $Q$ and also a significant 
	 variation of $C_1$ with the scale. The implications of this
	 scale dependence can be seen by choosing a scale $Q_0$ at
	 which the loop corrections to $\mu$ are minimized. At 
	 this scale one then finds that sign($C_1(Q_0)$)=-1.
	 One immediately sees that the implication of this result
	 is to change the nature of the radiative symmetry breaking
	 equation above from an ellipsoidal to a hyperbolic constraint.
	 The choice of $Q_0$ in the discussion above is for 
	 illustration purposes only and the phenomenon  discussed above 
	 would occur for any $Q_0$ in the region of the electroweak
	 symmetry breaking.	 
Now the parameter $\mu$  can also be regarded as the fine tuning 
parameter\cite{ccn} of the theory. Though there are by now  
several different criteria of what constitutes fine tuning\cite{finetuning}
the parameter $\mu$ provides the simplest criterion. This parameter
is especially suitable for interpreting the implications of 
the hyperbolic branch vs the ellipsoidal branch.
Thus for fixed $\mu$ and hence for a  fixed fine tuning, one finds
 that the ellipsoidal branch of radiative breaking 
 puts upper bounds on  
$m_0$ and $m_{1/2}$ while the hyperbolic branch does not. In the
latter case one finds that $m_0$ and $m_{1/2}$ could lie in the 
several TeV region consistent with a small $\mu$.

If indeed the 
hyperbolic branch is realized masses of some of the supersymmetric
particles, specifically the squarks, the gluino, the heavier 
neutralinos, the heavier chargino, and the heavy Higgs,
could be very large.  In this scenario the lightest particles
will be the light Higgs and $\chi_1^0, \chi_2^0, \chi_1^{\pm}$.
Typically for large $m_0$ and $m_{1/2}$ the pattern of 
masses for $\chi_1^0, \chi_2^0, \chi_1^{\pm}$ is given by 
   $m_{\chi_1^0}< m_{\chi_1^{\pm}}<m_{\chi_2^{0}}$
   at the tree level but  loop corrections here
   may be significant\cite{pierce,drees}. 
 In this scenario the mass differences 
 $\Delta M^{\pm} = m_{\chi_1^{\pm}}- m_{\chi_1^0}$
 and $\Delta M^{0} = m_{\chi_2^{0}}- m_{\chi_1^0}$ are typically
 O(10) GeV.  Thus the usual strategies for identification of
 supersymmetry in this region does not work and one must
 follow other strategies for identification of such
  particles\cite{cdg,fmrss,delphi}.  
  However, it was argued in Ref.\cite{ccnwmap}
  that observation of supersymmetric dark matter is still possible
  even in this region. This comes about  because as 
  $m_0$ and $m_{1/2}$ get large for fixed $\mu$,
$\chi_1^0, \chi_2^0$ and $\chi_1^{\pm}$  move
from being mostly gaugino like to mostly higgsino like.
  We will discuss more on this in the section on dark matter below. 
  A part of the hyperbolic region corresponds to the so called focus point
  region (FP)\cite{fmm}. As in the hyperbolic branch (HB) the focus point 
  region also corresponds to a small $\mu$. However, the focus point case
  corresponds to that part of the hyperbolic branch for which 
 $m_{1/2}$ is relatively  small.  As a result $m_0$ is also
 constrained to get not too large. Still  values of $m_0$ in several
 TeV region can be gotten in this part of the hyperbolic branch. 
 Thus the focus point region is contained in the hyperbolic branch
 and corresponds to the low end of the $m_{1/2}$ region on this
 branch (see also Refs.\cite{hb/fp,ccnwmap} in this context).
  Of course if the $g_{\mu}-2$ experimental
 difference at the level currently seen continues to persist then a 
 significant part of HB/FP region will be eliminated.

\section{Dark Matter in SUGRA}
Soon after the formulation of mSUGRA it was realized that the
lightest neutralino with R parity conservation could be a candidate for 
dark matter\cite{goldberg,Ellis:1983ew}. While this was
originally just a possibility a concrete realization of 
this possibility occurs when one carries out renormalization
group analyses on sparticle masses in mSUGRA models and one 
does indeed produce the light neutralino as the lsp over a 
significant part of the parameters space.
Further, as pointed out in Sec.2 in SUGRA models theoretical analyses show that  
for regions of the parameter space where $\mu >> M_Z$ 
($m_0,m_{\tilde g}\leq 1$ TeV), the lightest  
neutralino is a Bino\cite{scaling,roberts}. Of course, as discussed above
 there are  other
regions (HB/FP) where the lightest
neutralino would be mostly a higgsino. We discuss now some 
salient features of the analyses of supersymmetric dark matter.
First one needs to check if the density of relic neutralinos
falls within current limits given by the 
astrophysical observations.
The quantity of interest is $\Omega_{CDM}h^2$ where 
$\Omega_{CDM}=\rho_{CDM}/\rho_c$ where $\rho_{CDM}$ is the mass
density of cold dark matter in the universe and $\rho_c$ is the
relic density needed to close the universe, and $h$ is the
Hubble parameter in units of 100 km/sMpc. The most recent data from
the Wilkinson Microwave Anisotropy Probe indicates 
the result\cite{bennett,spergel} 
$\Omega_{CDM}h^2 = 0.1126^{+0.008}_{-0.009}$.
Before we discuss the implications of this highly accurate
determination for supersymmetry, we continue first with our 
general line of discussion. The analysis of the relic density
is quite intricate in that the annihilation of the relic
neutralinos can result in many final states, such $f\bar f$,
$WW$, $ZZ$, $Zh$ etc with the number of final states included
depending on the mass of the relic neutralino. The analysis of the
relic density involves the thermal averaged quantity 
$<\sigma_{eff} v>$ where $\sigma_{eff}$ is the neutralino 
annihilation cross  section and $v$ is the relative velocity
of the annihilating neutralinos.

In general thermal averaging will involve integrating over the 
Breit-Wigner poles which is somewhat of a delicate 
procedure\cite{greist,accurate}.
It turns out that the analysis of relic density is also significantly
affected by the phenomenon of 
coannihilation \cite{mizuta,efo1,efo2,ads,efo3,nihei,bednyk}. 
The quantity of interest in these analyses is
the number density $n=\sum n_a$ where the sum runs over all the
particle types that coannihilate and $n$ obeys the
Boltzmann equation 
$\frac{dn}{dt}$= $-3H n -<\sigma_{eff} v> (n^2-n_0^2)$ where
$H$ is the Hubble parameter and $n_0$ stands for the equilibrium
number density while $\sigma_{eff} =\sum\sigma_{ab} r_a r_b$ 
where $\sigma_{ab}$ is the annihilation cross section of particles
a and b and $r_a=n_{0a}/n_0$ with $n_{0a}$  the density  of
 particles of species $a$ at equilibrium. In the coannihilation 
 process after the freeze out the next to the lowest supersymmetric
 particles (nlsp's) decay to the lsp and thus n becomes the number
 density of the lsp. The importance of coannihilation arises
 from the fact that it extends considerably the allowed region of 
 the parameter space where the relic density constraints can 
 be satisfied. Thus without coannihilation the allowed range of the
 neutralino mass where the relic density constraint can be satisfied
 extends typically to around 150-200 GeV in mSUGRA. However, with
 the inclusion of coannihilation the allowed range of the 
 neutralino mass can 
 extend up to around 700 GeV\cite{efo2,ads}. In this  case the
 processes that enter in the relic density analyses are  
 $\chi \tilde \ell_R^a \rightarrow \ell^a \gamma, \ell^a Z, \ell^a h$,
$\tilde \ell_R^a \tilde \ell_R^b \rightarrow \ell^a \ell^b$,
 and $\tilde \ell_R^a \tilde \ell_R^{b*} \rightarrow \ell^a\bar \ell^b,
\gamma \gamma, \gamma Z, ZZ, W^+W^-, hh$. The most important 
coannihilation channel here turns out to be 
the one involving the staus.
 
Theoretical analyses over more than a decade
and a half have investigated both the indirect
and the direct detection of dark 
matter\cite{direct,sugradark,bottino1,baer2002} and a number 
of phenomena have been studied. These include the effects of 
nonuniversalities in the higgs sector and in the third generation
sector\cite{matallio,nonuni}. Specifically in these  sectors the
nonuniversalities can be parametrized at the GUT scale by the relations 
$m_{H_i}(M_G)=m_0(1+\delta_i)$ (i=1,2) and  by 
 $m_{\tilde t_L}(M_G)=m_0(1+\delta_3)$ and  
$m_{\tilde t_R}(M_G)=m_0(1+\delta_4)$ where one may 
allow $\delta_i$ (i=1-4) to be O(1)  consistent with the FCNC
constraints. It is found that variations of
$\delta_i$ in this range can lead to enhancements of the neutralino-proton
cross section ($\sigma_{\chi -p}$) by as much a factor of 10.
The gaugino non-universalities\cite{nonunigaugino,nonuniso10}
 also have a very significant 
influence on dark matter\cite{corsetti2,nelson}.
Nouniversalities in the gaugino sector can enter via non-singlet
representations in the decomposition of the neutralino mass matrix.
Thus, for example, for SU(5) the gaugino mass matrix transforms in
general like the symmetric product $24\times 24$ which in its 
decomposition contains the SU(5) representations 1,24, 75, and
200. The assumption of only the singlet leads to universal gaugino
masses while a non-vanishing contribution from the non-singlet
parts will lead to nonuniversalities. A similar situation
occurs for the case of SO(10) where the gaugino mass matrix transforms
like the symmetric product $45\times 45$ and contains the 
representations 1, 54, 210 and 770. The gaugino mass nonuniversalities
affect very significantly the allowed range of the neutralino 
mass over which the relic density constraints can be 
satisfied\cite{corsetti2}. Further, the direct detection rates are
also affected. Similarly, the effect of CP violation on dark
matter analyses turn out to be important\cite{cin}.
The situation regarding the imposition of the Yukawa unification
constraint and specifically the $b-\tau$ unification constraint
is interesting\cite{gomez,corsetti3}.  It is well known that $b-\tau$ 
unification requires a negative contribution to the b quark 
mass\cite{Pierce:1996zz,deBoer:2001xp}. Now roughly the correction to the b quark mass depends 
directly on the sign of $\mu$ and thus a negative contribution
to the b quark mass indicates that a negative sign of $\mu$ is 
preferable. However, the BNL experiment appears to indicate 
that the sign of $\mu$ is positive. Thus these two results appear
to be in  conflict. However, a closer scrutiny reveals that the
sign of the b quark correction depends on the sign $\mu m_{\tilde g}$
while the sign of $g_{\mu}-2$ correction is controlled by the 
the chargino exchange and hence depends on $\mu$ and the
$SU(2)$ gaugino mass $\tilde m_2$. Thus  one obvious solution
presents itself, i.e., that the signs of $m_{\tilde g}$ and 
$\tilde m_2$ are opposite\cite{Chattopadhyay:2001mj}.
 Specifically, choosing a negative 
sign for $\mu m_{\tilde g}$ resolves this conflict. The 
opposite correlation of $m_{\tilde g}$ and $\tilde m_2$ arises
naturally if the gauginos belong to the 24 plet representation for
the $SU(5)$ case and to the 54 plet representation for the
$SO(10)$ case.  The analysis of dark matter in this framework of Yukawa
unification is given in Ref.\cite{corsetti3}.
Other possibilities for this resolution have
also been explored\cite{Baer:2001yy,Tobe:2003bc}. 

We turn now to a discussion of dark matter on the hyperbolic 
branch. It is quite interesting that on the hyperbolc branch
we can satisfy the relic density constraints even though
much of the sparticle spectrum is rather heavy. The satisfaction
of the relic density constraints here arises once again 
due in part to the inclusion of coannihilation which arise
because of the near degeneracy of  $\chi_1^0$, ${\chi_2^0}$ and  
$\chi_1^{\pm}$.  The coannihilation involving these particles lead
to processes of the type 
$\chi_1^{+} \chi_1^{-}, \chi_1^0 \chi_2^{0}$$\rightarrow $
$u_i\bar u_i, d_i \bar d_i, W^+W^-$ and 
$\chi_1^0 \chi_1^{+}, \chi_2^0 \chi_1^{+}$
$\rightarrow $$u_i\bar d_i, \bar e_i\nu_i, AW^+,Z W^+, W^+h$.
The channel that dominates the coannihilation is the one that involves
 the sparticle which has the smallest mass
 difference with the lsp which in this case is between
$\chi_1^+$ and $\chi_1^{0}$.   As pointed out earlier the
neutralino in the hyperbolic region is mostly a higgsino 
and this structure tends to enhance the neutralino-proton 
cross section. 
We return now to the constraint of the WMAP 
constraint\cite{bennett,spergel}. There have been
several analyses recently to explore the implications  of
this constraint\cite{elliswmap,hb/fp,ccnwmap}.
  One important result that emerges
is that the new data limits more severely the parameters space
of models. Quite interestingly HB/FP region is consistent with the
WMAP constraint\cite{ccnwmap}. Further, the WMAP constraint produces
neutralino-proton cross sections that lie within range of the 
current\cite{dama,cdms,hdms,edelweiss} and  future\cite{genius,cline}
dark matter experiments\cite{ccnwmap}. 

\section{SUGRA, GUTS and Strings}
Supergravity grand unification much like SUSY GUTS generates baryon 
and lepton number violation whose nature and strength is 
 controlled by the nature of the grand unification group.
Theories based on SU(5), SO(10), E(6)  gauge groups generate below
the grand unification scale baryon and lepton number violating
dimension five operators with chiral structures 
LLLL and RRRR\cite{wein1,acn}.
When dressed with the full set of chargino, gluino and neutralino 
exchange diagrams the dimension five operators produce dimension
six  operators with chiral structures $LLLL$, $LLRR$, $RRLL$ 
and $RRRR$ which can decay the proton\cite{wein1,acn} and
a similar situation exists also in string models\cite{anstring}.
A detailed analysis of proton decay, however, is rather intricate
and depends on both the high and the low energy structures of the theory.
Thus the lifetime of the proton
can be significantly affected by the soft breaking sector of
SUGRA GUTS and by  the Higgs triplet 
structure\cite{multi} and specifically by the textures in the Higgs triplet
sector\cite{Nath:1996qs}. The most recent limits on the proton lifetime 
appear to disfavor the minimal SU(5) model\cite{raby}
(see, however, Ref.\cite{bajc}).
Regarding SO(10) there are a whole variety
of possible SO(10) models and so there are no necessarily definitive SO(10)
predictions since the proton decay modes are highly model
dependent\cite{so10a,so10b,so10c,so10d}. An important issue 
concerns the role of large representations such as 120 and 
$\overline{126}$. The appearance of such representations can 
significantly affect analyses of 
proton decay\cite{so10d} and of neutrino masses\cite{mahanthappa}.
We turn now to a brief discussion of the connection of SUGRA models 
and strings. Since SUGRA  models are derived from models involving
supergravity and supergravity may be viewed as a low energy limit
of string theory below the Planck scale, it is natural to imagine
 SUGRA arising as a low energy limit of a string model.
There are two elements involved  in such a connection. First one 
must try to deduce a realistic model with a standard model gauge
group from string theory and efforts have been in this direction
from the very beginning\cite{heterotic,stringuts}. Second one
must try to obtain a realistic breaking of supersymmetry from
strings and there has some been progress also along these
lines\cite{nilles1,nilles} specifically using dualities\cite{fmtv,font}.
More recently the constraints of modular invariance on soft 
breaking have been investigated to make contact with
low energy phenomenology\cite{gaillard}. An interesting 
issue concerns the constraints needed in modular invariant
theories to derive universality of soft parameters
and in Ref.\cite{nathtaylor} some dynamical constraints to achieve
universality of soft parameters were identified. 
Thus while we do not yet have a fully realistic string model 
it is interesting that one can still make  tentative contact
between supergravity based models and string theory.

\section{Conclusion}
The advent of SUGRA models in 1982 spurred an activity in 
supersymmetry phenomenology that still continues. Historically 
it was only within the framework of SUGRA models that a 
phenomenologically consistent spontaneous  
breaking of supersymmetry was first achieved. The basic concept
of supersymmetry breaking in one sector and its communication to
the physical sector also introduced first in SUGRA finds applications
in string based scenarios. Further, SUGRA models with R parity predict 
the existence of cold dark matter (CDM), something that appears
desirable from astrophysical considerations.
The literature on mSUGRA, its extensions  and their implications 
is enormous and a comprehensive review of the developments 
is obviously outside the scope of a conference talk. 
Thus we have focussed on a few topics of current interest.
 One of these topics concerns the difference
$\Delta a_{\mu}= (a_{\mu}^{exp}-a_{\mu}^{SM})$. There are several 
estimates of this quantity which differ mainly due to the different
estimates of the leading order hadronic correction. According to
the recent analysis of Ref.\cite{hagiwara} this difference is
$(33\pm 10)\times 10^{-10}$ which amounts to about a $3\sigma$ deviation
between experiment and theory. An effect of this size is expected 
within SUGRA models since it was noted early on\cite{yuan}
that the size of the supersymmetric electroweak correction could
be as large or larger than the standard model electroweak correction.
However, the theoretical evaluations of the hadronic error are still
in a state of flux and $\Delta a_{\mu}$ is likely to shift before
it settles down. However, we note that  if  a value of  $\Delta a_{\mu}$ 
 persists at a perceptible 
 level, i.e.,  $\sim  10^{-9}$ then the sparticle mass limits 
 lie within reach of the LHC and the 
 direct observation of new physics is implied. Thus most of the sparticles 
 ($\tilde g, \tilde q, \tilde W, \chi^0,..$) should  become visible at the LHC.
Another interesting aspect of SUGRA models is that renormalization
group analyses show that in a large part of the parameter space
the sign of $a_{\mu}^{SUSY}$ is correlated with the sign of 
$\mu$. Thus the current analyses on $\Delta a_{\mu}$ 
imply a positivity of the $\mu$ sign.  
A positive $\mu$ is very desirable for the satisfaction of the
  $b\rightarrow s+\gamma$ constraint and also for the observation
  of supersymmetric dark matter. 
  
  mSUGRA is consistent with the flavor changing neutral current
  constraints. However, it is possible to extend mSUGRA to 
  include non-universalities by the assumption of a non-flat 
  Kahler potential and  a non-flat gauge kinetic energy function.
   These extensions allow one to include nonuniversalities in the
  Higgs sector, in the third generation sector and in the 
  gaugino masses consistent with  the FCNC constraints.
  Further, SUGRA models allow the soft breaking parameters to
  become complex in general. Thus mSUGRA allows up to two phases
  in the soft breaking sector while more phases can appear when
  one includes nonuniversalities. However, the inclusion of 
  phases requires strong consistency checks with the
  current very sensitive limits on the electron and the neutron
  edms. Additionally, the atomic edms generated via the Schiff
  moments also constrain phases. This is the case specifically
  for the atomic edm of $Hg^{199}$. Typically phases O(1)
  will violate these constraints unless a mechanism is invoked
  for their suppression. Several mechanisms for such 
  suppressions have been discussed in the literature. One mechanism
  which leads to a natural suppression in certain regions of the
  parameter space of SUGRA models is the cancellation mechanism 
  and there are many works exploiting this technique to allow
  for large phases. However, whatever mechanism is employed for
  the suppression of the edms, the presence of large phases typically
  has large effects on supersymmetry phenomenology. One of the
  most dramatic effects occurs in the Higgs sector where the
  Higgs mass eigenstates are no longer CP even and CP odd states 
  but rather admixtures of CP even and CP odd states. 
  This mixing will lead to rather dramatic effects in SUSY 
  phenomena at   $e^+e^-$
  colliders and elsewhere. Further, the inclusion of phases 
  produces important effects on SUSY corrections to the Higgs mass,
  on SUSY corrections to the b quark and $\tau$ lepton masses and 
  on Higgs decays to $b\bar b$, $\tau\bar \tau$ and $c\bar c$.
  Thus, for example, the  decay branching ratio of the Higgs 
  to $b\bar b$ will carry signatures of both supersymmetry and CP phases.
  Another process affected strongly by phases is the 
  decay  $B^0_{s,d}\rightarrow \mu^+\mu^-$.
  The branching ratio for this process in the standard model
  is too small to be accessible to experiment in RUNII of the
  Tevatron. However, in SUGRA models the branching ratio for
  this process can be enhanced by as much as a factor of $10^2$ 
  for large $\tan\beta$. Further, the inclusion of CP phases
  can produce additional enhancements which can be as large  
  as another factor of $10^2$. These enhancements put the 
  the $B^0_{s,d}\rightarrow \mu^+\mu^-$ branching ratio within
  reach of the Tevatron. Thus the observation of this process
  will be a strong hint for supersymmetry pointing to the
  existence of sparticles even before the sparticles are 
  directly observed.
  
  The existence of supersymmetric dark matter is an important
  prediction of SUGRA models with R parity invariance. 
  Detailed analyses of the density of the relic neutralinos
  indicate that the predictions of mSUGRA and its extensions
  allow for consistency with the most recent determinations
  of $\Omega_{CDM}h^2$ from the WMAP data. Further,
  the relic density limits from WMAP more sharply constrain
  the sparticle spectrum and define more sharply the allowed
  ranges of the spin independent and spin dependent neutralino-
  proton cross sections. In this talk we have also reviewed
  the hyperbolic branch of the radiative breaking of the
  electroweak symmetry. A part of this branch allows for
  large values of $m_0$  and $m_{1/2}$ for a fixed 
  value of $\mu$ and puts most of the sparticle in this region
  in the several TeV region. However, quite interestingly 
  this region  still produces a relic density consistent 
  with the WMAP constraints and leads to scalar and spin dependent
  neutralino- proton cross sections which appear to be within
  reach of the future dark matter experiments such as GENIUS and ZEPLIN.
  Proton decay in SUGRA GUTs depends on two elements, on the 
  sparticle spectrum and on the GUT group. Unlike the sparticle
  masses which in mSUGRA are essentially independent of the GUT structure
  as pointed out in section 1, proton decay hinges critically on
  the GUT structure. Thus this sector of the theory is more
  model dependent. While models do exist where one can make
  consistent the GUT theory with the current proton decay limits,
  there is not a uniqueness in fixing the GUT structure.
 Fortunately, the low energy predictions of SUGRA models are 
 independent of the dimension 5 operators and thus the
 SUGRA predictions are not affected by issues related to
 the proton lifetime. Finally, we note that
SUGRA models have gravity as an intrinsic piece of their
fabric and have good chance of making contact with string theory.
Thus  more effort is needed to derive SUGRA, mSUGRA  and other 
competing models from  a top down approach. 
More than 20 Years after its invention SUGRA is still
a leading candidate  for new physics beyond the SM. 
Experiment is awaited to check the predictions of this model
in the laboratory.\\

\noindent
{\bf Acknowledgments}\\ 
  This research was supported in part by NSF grant PHY-0139967.
 This article was written when the author was visiting the 
 Max Planck Institute fur Kernphysik, Heidelberg.
  The author thanks the Institute for hospitality and  the Alexander 
  von Humboldt Foundation for support during this period.

\end{document}